# Aromatic borozene

N. Gonzalez Szwacki,[1] V. Weber,[2] and C. J. Tymczak[1]

[1]*Department of Physics, Texas Southern University, Houston, Texas 77004, USA*
[2]*Department of Physical Chemistry, University of Zurich, 8057 Zurich, Switzerland*

Based on our comprehensive theoretical investigation and known experimental results for small boron clusters, we predict the existence of a novel aromatic inorganic molecule, $B_{12}H_6$. This molecule, which we refer to as *borozene*, has remarkably similar properties to the well-known benzene. Borozene is planar, possesses a large first excitation energy, $D_{3h}$ symmetry, and more importantly is aromatic. Furthermore, the calculated anisotropy of the magnetic susceptibility of borozene is three times larger in absolute value than for benzene. Finally, we can show that borozene molecules may be fused together to give larger aromatic compounds with even larger anisotropic susceptibilities.

## I. INTRODUCTION

Why certain molecules are more stable than others is not always easy to understand. Nature's diversity does not always permit a simple answer for the structure of all compounds. A very useful concept in structural stability is aromaticity,[1] which was first developed to account for the properties of organic compounds involving ring structures such as benzene ($C_6H_6$) and more recently extended to inorganic systems.[2] However, the question arises if aromatic hydrocarbons are the only structures where an "aromatic ring" acts as a building block and plays a key role in their stability. In this study we predict the existence of an inorganic molecule, $B_{12}H_6$, which has remarkably similar properties to benzene. This molecule, which we call *borozene*, is planar, possesses a large first excitation energy, exhibits a highly aromatic character, and similar to benzene is a building block of much larger aromatic compounds.

Small all-boron clusters, $B_n$ ($n < 20$) have been widely investigated both experimentally and theoretically.[3-5] All these studies indicate that small boron clusters assume in most cases quasi-planar structures, and in some special cases even perfectly planar structures. In contrast, neutral and anionic boron hydrides, $B_nH_{n+m}$, are all known to have three-dimensional deltahedral structures.[6] There is yet little known about the structure of small boron hydrides where the number of hydrogen atoms is smaller than the number of boron atoms (see Ref. 7 and references therein). One such example is the recently studied σ-aromatic and π-anti-aromatic $B_7H_2^-$ cluster which is fully planar.[7]

It was experimentally established that one of the most stable all-boron clusters is made up of twelve boron atoms, is quasi-planar in shape, and possesses a large first excitation energy of 2.0 eV.[3] The $B_{12}$ structure consists of 13 $B_3$ triangles with 12 outer triangles surrounding a central one; the atoms forming the central triangle are situated above a nine-member boron ring making $B_{12}$ a convex structure of $C_{3v}$ symmetry.[3,4] Our calculations revealed that the $B_{12}$ cluster has three outer boron pairs that are 5% shorter than the average B−B bond lengths between the rest of the boron atoms. This suggests the presence of strong covalent bonds between those atoms. However, it is possible to increase the B−B bond lengths of the three outer pairs by 17% by attaching hydrogen atoms to the outer boron atoms (see Figure 1a). As a consequence, the molecule becomes perfectly planar. This finding motivated us to investigate the interaction between $B_{12}$ cluster and up to four hydrogen molecules.

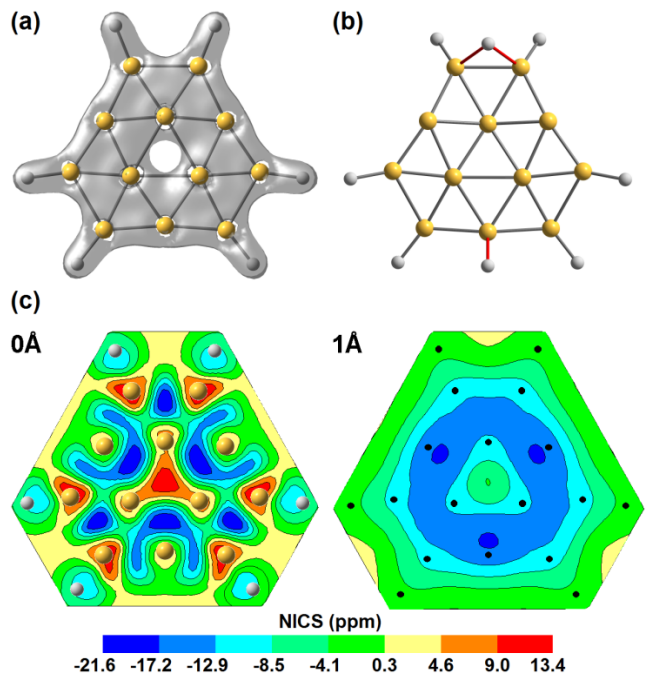

FIG. 1. (a) Plot of the structures and total electronic densities of $B_{12}H_6$. Note that the density of electrons is weaker at the center of the "boron ring." (b) Quasi-planar $C_s$ structure of $B_{12}H_8$, which is the energetically preferred configuration for $B_{12}$ with 4 $H_2$ molecules, attached to it. (c) Contour plot of NICS(x, y) for $B_{12}H_6$ in plane (left) and at 1 Å above the planar molecule (right).

## II. COMPUTATIONAL DETAILS

The structure and electronic properties of all clusters were obtained at the X3LYP/6-311G(d,p) level of theory using tight convergence criteria as implemented in MondoSCF, a suite of programs using Gaussian basis sets and all-electron Hartree-Fock, density functional theory or hybrid approach for self-consistent electronic structure calculations.[8] The initial search for the most stable structures of the boron hydride $B_{12}H_n$ have been done at the X3LYP/6-31G(d,p) level of theory starting from the energy-minimum structure of $B_{12}H_{n-2}$ and the low-lying isomers in each case have been re-optimized using the 6-311G(d,p) bases set. The obtained energy-minimum structures are well separated in energy from its higher isomers by at least 22 kcal/mol in the case of $B_{12}H_n$, where $n \leq 6$, and 17 kcal/mol in the case of the $B_{12}H_8$ cluster. The $B_{22}H_8$ and $B_{60}H_{12}$ clusters, which result from the fusion of two and six $B_{12}H_6$ molecules, respectively, were fully optimized using symmetry-unrestricted calculations. To ensure that the planar structures correspond to a minimum of energy the nature of the stationary points has been checked by vibrational frequency calculation. The HOMO-LUMO gap is defined as the energy separation between the highest occupied molecular orbital (HOMO) and the lowest unoccupied molecular orbital (LUMO).

The first singlet excitation energy, Nuclear Magnetic Resonance (NMR) shielding tensors, and magnetic susceptibility tensors were calculated using the Gaussian03 package.[9] To obtain the NICS values (from the NMR shielding tensors) we have used the GIAO (Gauge-Independent Atomic Orbital) method and the magnetic susceptibility tensors were calculated using the CSGT (Continuous Set of Gauge Transformations) method. All computations have been performed at the X3LYP/6-311++G(d,p) level of theory except for the $B_{60}H_{12}$ cluster for which we have used the RHF/6-31G(d,p) level of theory. The anisotropy of magnetic susceptibility is defined as the difference between out-of-plane and the average in-plane components of the susceptibility tensor.

The MOs of $B_{12}H_6$ and $C_6H_6$ were calculated at the RHF/6-311++G(d,p) level of theory using the GAMESS-US package.[10] The same package was used to calculate the π-π interaction between molecules in borozene and benzene dimers at the RHF-MP2/6-311G(d,p) and RHF-MP2/6-311++G(d,p) levels of theory, respectively. The counterpoise correction was applied to account for the basis set superposition error.

## III. RESULTS AND DISCUTION

The search for the most stable structures of $B_{12}H_n$, with $n \leq 6$ is simplified by the fact that the most energetically favorable configurations are those where the hydrogen atoms are directly attached to the outer boron atoms of the molecule. We have established that the most likely stable configuration for $B_{12}H_2$ is when the hydrogen atoms are attached to one of the outer short-bonded boron pairs of the $B_{12}$ cluster. The energetically preferred configuration for $B_{12}H_4$ is when the hydrogen atoms are attached to one of the two remaining outer short-bonded boron pairs in the $B_{12}H_2$. Finally, the $B_{12}H_6$ cluster has all short B–B pairs, from $B_{12}$, with hydrogen atoms attached to them. Only $B_{12}H_6$ is a fully planar molecule, whereas $B_{12}H_2$ and $B_{12}H_4$ are quasi-planar with $C_s$ symmetry. In Figure 1a we have shown the structure of $B_{12}H_6$. The hydrogenation energy defined as $\Delta E = E(B_{12}H_n) - E(B_{12}H_{n-2}) - E(H_2)$, where $E$ is the total energy, is -44 kcal/mol for $n= 2$, -45 kcal/mol for $n= 4$, and -51 kcal/mol for $n= 6$. We have found, however, that if a fourth $H_2$ molecule is attached to $B_{12}H_6$ the hydrogenation energy increases to -2 kcal/mol (i.e. the $H_2$ molecule is weakly bound to $B_{12}H_6$). It is also important to mention that our $\Delta E$ values are about two times larger than the predicted energy of hydrogenation of the $B_7^-$ cluster,[7] which is an indication of unusual stability of the $B_{12}$ structure. The $B_{12}H_8$ molecule is shown in Figure 1b and can be described as a distorted $B_{12}H_6$ cluster with two extra (one terminal and one bridging) hydrogen atoms attached to it. The B–H bond lengths are 1.36 and 1.21Å for the bridging and terminal hydrogen atoms, respectively, whereas the remaining B–H distances in $B_{12}H_8$ and in all other described above $B_{12}H_n$ ($n \leq 6$) clusters are the same and equal to 1.18 Å. The last value is very close to the calculated bond lengths B–H= 1.19Å in borane, $BH_3$.

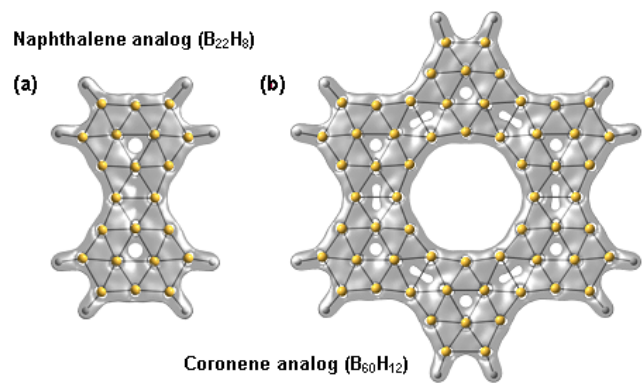

FIG. 2. Plot of the structures and total electronic densities of (a) $B_{22}H_8$, and (b) $B_{60}H_{12}$. In both molecules all B–H distances are the same and equal to 1.18 Å.

Although no single measure of aromaticity is without limitations, the anisotropy of the magnetic susceptibility (AMS) is an important indicator of diatropicity.[11] The $B_{12}H_6$ molecule has very important properties: it is planar with $D_{3h}$ symmetry; it possesses large first excitation energy of 2.6 eV and large AMS of -208.2 cgs-ppm. Also, the $B_{12}H_6$ molecule can be a building block of larger planar molecules with similar structural and physical characteristics. In Figure 2a and 2b are shown what we call boron analogues of naphthalene ($B_{22}H_8$) and coronene ($B_{60}H_{12}$), which are fusions of two and six $B_{12}H_6$, clusters, respectively. Also of interest, the HOMO-LUMO gap decreases with cluster size; the gap values are 3.6, 2.4, 1.3 eV for $B_{12}H_6$, $B_{22}H_8$, $B_{60}H_{12}$, respectively. Furthermore, the absolute AMS value increases with cluster size; the AMS values are -294.3, -454.7 for $B_{22}H_8$, $B_{60}H_{12}$, respectively. This and other values for boron hydrides have been summarized in Table 1. In this table we have also included the calculated values for $B_{12}$ with enforced planarity ($D_{3h}$) and fully relaxed ($C_{3v}$) symmetries. We have in addition included the values for three hydrocarbons for comparison. It should be noted that the absolute value of AMS for $B_{12}H_6$ is three times larger than our value for benzene (-67.5 cgs-ppm) and 7% larger than the value for the $C_{3v}$ $B_{12}$ cluster (-192.9 cgs-ppm).

TABLE 1. Molecular Symmetry, HOMO-LUMO energy gaps, and the isotropic and anisotropic values of magnetic susceptibility for the studied planar boranes and hydrocarbons. For comparison we have also included our results for $B_{12}$ with enforced planarity ($D_{3h}$) and fully relaxed ($C_{3v}$).

| Structure | Symmetry | HOMO-LUMO gap (eV) | Magnetic susceptibility (cgs-ppm) | |
|---|---|---|---|---|
| | | | Isotropy | Anisotropy |
| $B_{12}$ | $C_{3v}$ | 3.73 | -105.4 | -192.9 |
| $B_{12}$ | $D_{3h}$ | 3.58 | -107.6 | -213.6 |
| $B_{12}H_6$ | $D_{3h}$ | 3.67 | -92.0 | -208.2 |
| $B_{22}H_8$ | $D_{2h}$ | 2.38 | -147.9 | -294.3 |
| $B_{60}H_{12}$ | $D_{6h}$ | 1.30 | -286.5 | -454.7 |
| $C_6H_6$ | $D_{6h}$ | 6.86 | -53.0 | -67.5 |
| $C_{10}H_8$ | $D_{2h}$ | 4.93 | -90.4 | -128.5 |
| $C_{24}H_{12}$ | $D_{6h}$ | 4.13 | -251.1 | -474.5 |

To gain information about the individual contributions of the $B_3$ triangles to the overall aromaticity of the $B_{12}H_6$ molecule we have studied its two-dimensional Nucleus Independent Chemical Shift (NICS) map. In Figure 1c we have shown the contour plot of NICS(x, y) in plane (left) and at 1 Å above the $B_{12}H_6$ molecule (right). It is clearly seen from the left part of the figure that the NICS values are negative inside the twelve outer $B_3$ triangles of the molecule, suggesting a flow of a global diatropic current around the central triangle. The central part of the molecule has a paratropic current flowing inside the inner $B_3$ triangle which is not overwhelmed by a diatropic current due to an electron charge transfer from the center of the structure towards the outer boron triangles after hydrogenation. Also, this anti-aromatic region is spatially localized since the NICS values are negative at 1 Å (see the right part of Figure 1c) and 2 Å above and below the center of the $B_{12}H_6$ molecule (NICS(1)= -3.9 ppm; NICS(2)= -5.2 ppm).

The benzene dimer is the simplest prototype of the aromatic π-π interactions which is an important weak interaction present in aromatic supra-molecular systems.[12] Since the number of π electrons in borozene and benzene are the same and the molecular orbital (MO) picture for these electors is very similar (see Figure 3) we may expect that the strength of the aromatic-aromatic interaction in a borozene dimer is comparable to that of the benzene dimer. To investigate this we have considered the simplest case where the molecules in the dimer have the parallel "sandwich" configuration. In Figure 4 we plotted the association energy versus the distance between the molecules in the $B_{12}H_6$ and $C_6H_6$ dimmers. From this figure we can see that the association energy for the borozene dimer, in its equilibrium position, is about five times larger than the corresponding energy for the benzene dimer. This result suggests a stronger polarization contribution from borozenes π-MOs, which we theorize is a consequence of more delocalized π-electrons in the borozene dimer in respect to the benzene dimer.

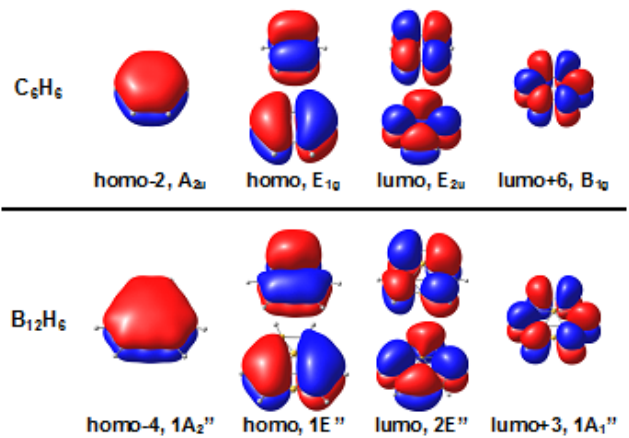

FIG. 3. Comparison of the π molecular orbitals of benzene with the corresponding π molecular orbitals of $B_{12}H_6$.

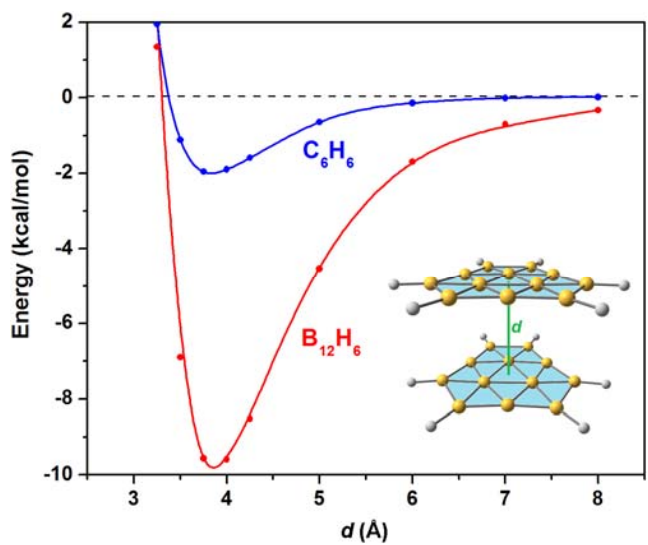

FIG. 4. Potential energy curves for $B_{12}H_6$ and benzene dimers versus the center-to-center distance between the monomers. The association energies are -1.99 and -9.81 kcal/mol and the equilibrium distances are 3.8 and 3.9 Å for $C_6H_6$ and $B_{12}H_6$, respectively.

Although the specific route for the synthesis of the $B_{12}H_6$ structure is not yet known, it is clear from our investigation that to some extent the chemistries of $B_{12}H_6$ and benzene may be very similar, suggesting that similar methods could be employed to synthesize this and related compounds. Given the technological importance of benzene and its derivatives, we believe that this molecule will have a significant technological impact and deserves further extensive study.

## ACKNOWLEDGMENTS

We would like to thank Dr. Daniel Vrinceanu for his helpful discussion and Brad Mazock for computational support. This project was supported by the Robert A. Welch Foundation (Grant J-1675) and the NIH-RCMI Program (Grant RR03045).